\documentclass[aps,prl,twocolumn,showpacs,groupedaddress]{revtex4}
\usepackage{amsfonts}
\usepackage{amsmath}
\usepackage{amssymb}
\usepackage{graphicx}
\begin{document}
\title{Indication for macroscopic quantum tunneling below ${\rm 10 \ K}$ in nanostructures of ${\rm SrRuO_3}$}
\author{Omer Sinwani$^{1}$}
\author{James W. Reiner$^{2}$}
\author{Lior Klein$^{1}$}
\affiliation{$^1$Department of Physics, Nano-magnetism Research
Center, Institute of Nanotechnology and Advanced Materials,
Bar-Ilan University,Ramat-Gan 52900, Israel}
\affiliation{$^{2}$Hitachi Global Storage Technologies, San Jose, California 95135 USA}

\keywords{}%

\begin{abstract}
We study magnetization reversal of nanostructures of the itinerant ferromagnet ${\rm SrRuO_3}$ ($T_c{\rm \sim 150 \ K}$).  We find
that down to ${\rm 10 \ K}$ the magnetization reversal is dominated by thermal activation. From ${\rm 2-\rm 10 \ K}$, the magnetization reversal becomes independent of temperature, raising the possibility for reversal dominated by macroscopic quantum tunneling (MQT). A ${\rm 10 \ K}$ crossover to MQT is consistent with the extremely large anisotropy field (${\rm \sim 7 \ T}$) of ${\rm SrRuO_3}$.

\end{abstract}

\maketitle

Quantum tunneling through a potential barrier is one of the most remarkable manifestations of quantum behavior. While there is good understanding and many experimental realizations of this phenomenon when the tunneling object is microscopic, the extension of this behavior to macroscopic objects poses one of the most intriguing theoretical and experimental challenges. A promising route, which we adopt here, is to look for signatures of MQT in magnetization reversal of ferromagnetic nanoparticles.

At elevated temperatures, the
magnetization reversal of ferromagnetic nanoparticles is
 commonly described in the framework of the N\'{e}el-Brown model \cite{Neel,Brown_pr,Brown_euro}.
In its simplest form, the model describes a thermally activated process of coherent rotation at a temperature $T$ over an energy barrier $E_b$, and it predicts an average waiting time $\tau$ given by  $\tau= \tau_0 e^{E_b/k_B T}$, where $\tau_0$ is a sample specific constant linked to Larmor frequency with a typical value around $10^{-9}$ s \cite{Cullity}.  However,  in the low temperature limit, a crossover is theoretically expected from a thermally activated reversal to a temperature-independent magnetization  reversal dominated by
 MQT \cite{MQT_theory_BOOK,MQT_theory}.

A low-temperature crossover to MQT-dominated reversal has been demonstrated using the magnetic molecules ${\rm Mn_{12}}$  \cite{Mn}and ${\rm [Fe_{8}O_{2}(OH)_{12}(tacn)_{6}]^{8+}}$  \cite{Fe}, both  with a spin ground state of $S=10$ with crossover temperatures of ${\rm 0.35}$ and ${\rm 0.6 \ K}$, respectively. The crossover was manifested in temperature independent hysteresis loops with a series of steps separated by plateaus indicating resonant tunneling.

Consisting of at least hundreds of spins, nanoparticles have an energy level spacing which is too small to be identified by resonant tunneling in hysteresis loops. Thus reports on MQT of nanoparticles are mainly based on the identification of a crossover from a thermally activated reversal to a temperature-independent reversal. These reports include temperature independence below ${\rm 5 \ K}$ of switching field distribution of  nickel nanowires  \cite{Ni} and temperature independence below ${\rm 0.35 \ K}$ of  two level fluctuations of self-assembled ErAs quantum wires and dots in semi-insulating GaAs matrix \cite{ErAs}. In addition, it has been shown that the magnetization reversal of ${\rm BaFe_{12-2x}Co_xTi_xO_{19}}$  deviates  from N\'{e}el-Brown model below ${\rm 0.4 \ K}$ \cite{BaFe} and the crossover temperature depends on the direction of the reversing field, in agreement with theoretical predictions  for MQT.

Here we  study patterned nanostructures of thin films of an extremely hard ferromagnet, ${\rm SrRuO_3}$, and show that when a reversing field that does not yield immediate reversal is applied above $\sim 10 \ {\rm K}$, the average waiting time for reversal increases sharply with decreasing temperature as expected for thermally activated reversal. On the other hand,  below  $\sim 10 \ {\rm K}$ the average waiting time is temperature independent. The results strongly suggest a crossover to MQT at a relatively high crossover temperature which is consistent with the high anisotropy field of ${\rm SrRuO_3}$. Furthermore, the results open new and exciting opportunities for elucidating the intriguing phenomenon of MQT.

For this study we use high quality epitaxial thin films of the itinerant ferromagnet ${\rm SrRuO_3}$  ( $T_c{\rm \sim 150 \ K}$) \cite{srruo} grown on a slightly miscut ${\rm SrTiO_3}$ substrate ($\sim 2^\circ$) by molecular beam epitaxy (MBE). The films are orthorhombic with lattice parameters $a ={\rm 5.53\ \AA}$, $b={\rm 5.57\ \AA}$ and $c={\rm 7.82\ \AA}$ and they grow with the $c$ axis in the film plane and the $a$ and $b$ axes at $45^\circ$ relative to the film normal \cite{Ms}.
The films have large uniaxial magnetocrystalline anisotropy where the anisotropy field is ${\rm \sim 7 \ T}$ \cite{Kerr} and the easy axis is
in the (001) plane. Above $T_c$ the easy axis is along $b$ \cite{EA_Kats} and below $T_c$ there is a reorientation transition  and the direction of the easy axis changes in the $(001)$ plane towards the film normal at a  practically constant rate of $~0.1^\circ$ per degree \cite{Ms}.   When the films are zero-field cooled, a stripe-domain structure emerges with domain walls parallel to the in-plane projection of the easy axis. The width of the magnetic domains is ${\rm \sim 200 \ nm}$ \cite{Domain strips} and the estimated wall width is $\sim3$ nm \cite{wall_width}.

\begin{figure}[t]
\includegraphics[scale=0.47]{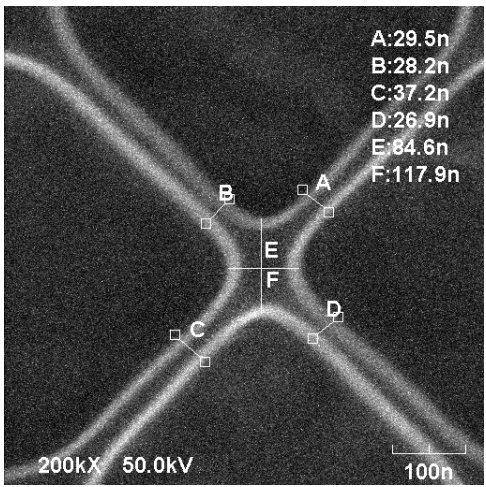}
\caption{A scanning electron microscope image of a typical pattern of $\rm{ SrRuO_{3}}$.}\label{trapping}
\end{figure}

Figure 1 shows a typical pattern of a ${\rm 7 \ nm}$ thick film used for this study. It consists of an internal rectangle ${\rm 115\pm10 \ nm \times 85\pm10 \ nm}$ connected by four narrow leads which are $ {\rm 30\pm10 \ nm}$ wide. The internal square and the leads are both made of ${\rm SrRuO_3}$. The patterns are fabricated
using a CABL-9000C e-beam high resolution lithography system (CRESTEC) followed by ${\rm Ar^+}$  ion milling.
The average magnetization in the internal square of the patterns is monitored by measuring the anomalous Hall effect (AHE), which is proportional to the average film-perpendicular component of the magnetization and therefore is commonly used for probing the magnetization  in patterned films. The measurements are performed using a PPMS system (Quantum Design) integrated with external electronics.
The setup allows the separation of symmetric and antisymmetric contributions by exchanging current and voltage leads \cite{reciprocity_theorem}.

 To identify a crossover to a temperature-independent reversal, it is important to use a sample with a well-characterized thermally activated regime. The simplest realization is a single domain particle with only two stable states separated by a single energy barrier. In such a case only full magnetization reversal is possible and good agreement with the N\'{e}el-Brown model is expected. Good agreement can also be expected in cases where the magnetization reversal occurs in two stages: first a small volume with reversed magnetization nucleates and then it propagates until full reversal is achieved. If the nucleation field is larger than the field required for full propagation, every nucleation leads to full reversal and if the energy barriers for nucleation are narrowly distributed a good agreement with the N\'{e}el-Brown model is expected. The probability of realizing both cases increases as the pattern dimension is decreased.

 Small size patterns (on the order of $\rm 100 \ nm \times 100 \ nm$) are usually fully magnetized when they are zero-field cooled (namely no domains are formed) and they also usually exhibit a superparamagnetic phase between $T_c$ and a blocking temperature ${T_{b}\sim116}$ K manifested in  zero-field spontaneous magnetization reversals. This intriguing regime will be addressed in detail elsewhere.
\begin{figure}[t]
\includegraphics[scale=0.47]{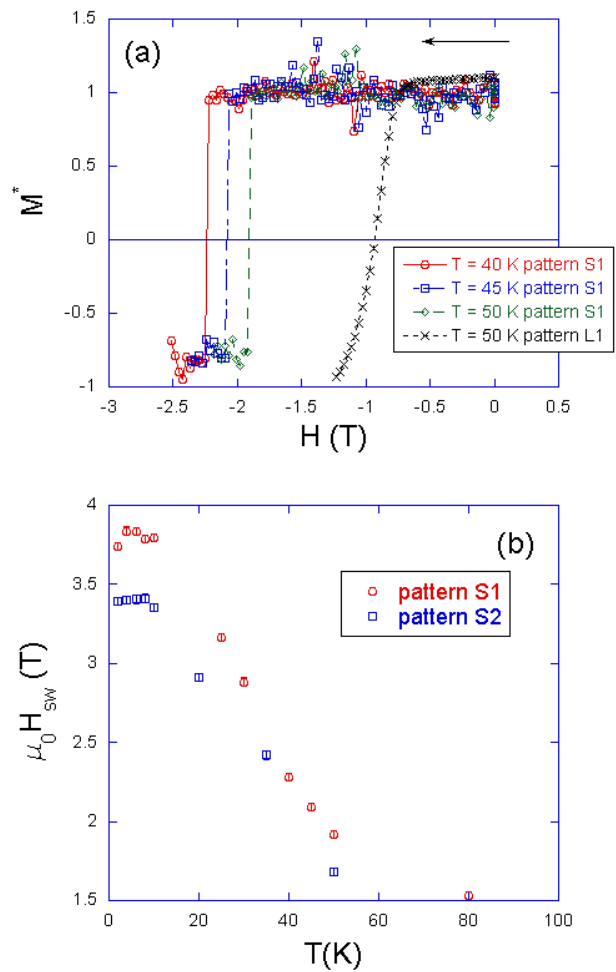}
\caption{(a) Normalized magnetization $\rm{M}^{\ast}$ as a function of a magnetic field $H$ applied at $60^{0}$ relative to the sample normal in the (001) plane and $\sim30^{0}$  relative to the easy axis. The magnetic field is swept  at a rate of  100 Oe/s and the magnetization is measured every $\sim3$ s. The internal rectangle of pattern S1 is ${\rm 115\pm10 \ nm \times 85\pm10 \ nm}$. The internal square of pattern L1 is $\sim{\rm 20 \ \mu m \times 20 \ \mu m}$. (b) The switching  field, determined in sweeping field experiments, as a function of temperature for different samples. Pattern S2 is similar in its dimension to pattern S1.} \label{trapping}
\end{figure}

 Below the blocking temperature, magnetization reversal is induced by field. Figure 2(a) shows reversals when the applied magnetic field is swept at a constant rate. We note the sharp full reversal of the patterns on the order of ${\rm 100 \ nm \times 100 \ nm}$ in contrast with the smooth and gradual reversal when the pattern is one hundred times larger. Figure 2(b) shows for each temperature the result of a single sweeping field experiment. We note a flattening of the switching field  below 10 K which motivates us to more closely examine the possibility of a crossover to temperature-independent magnetization  reversal.

\begin{figure}[p]
\includegraphics[scale=0.625]{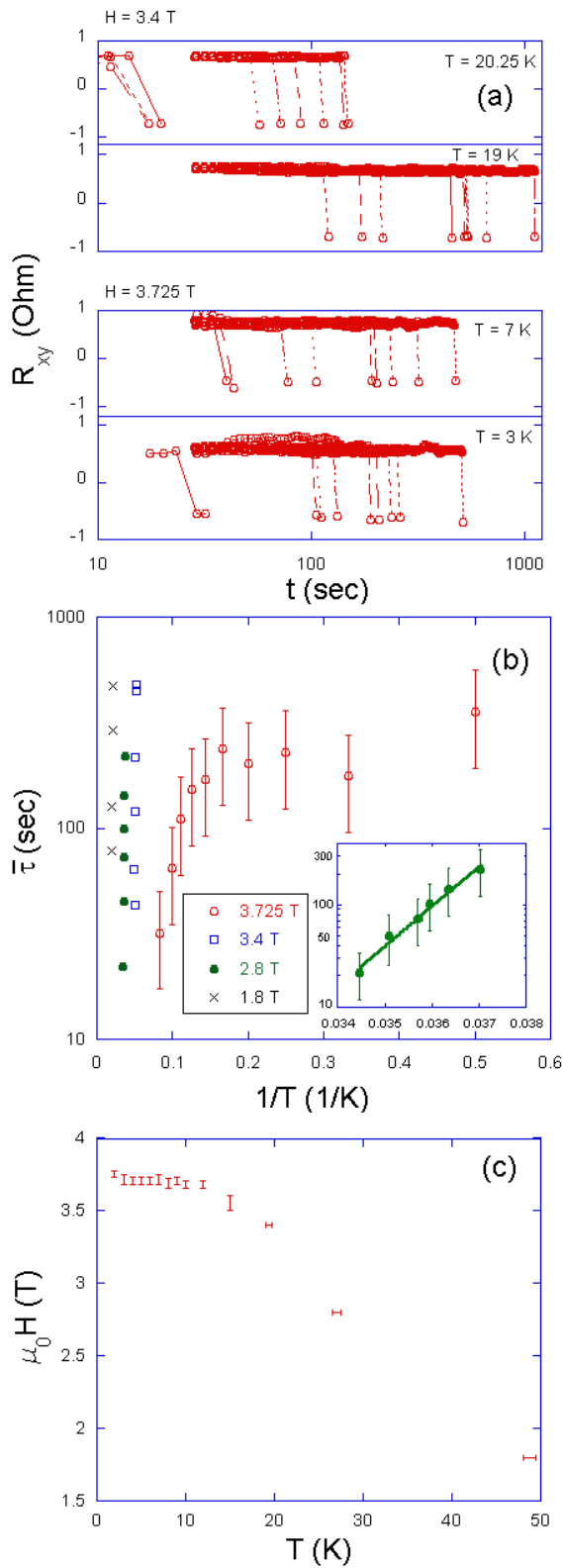}
\caption{(a) The transverse resistance $\rm{R_{xy}}$ as a function of time at different temperatures. At 19 K and 20.25 K,  $\mu_{0}H$= 3.4 T, at 3 K and 14 K $\mu_{0}H$=3.725 T. $H$ is applied at $\rm{60^{0}}$ relative to the sample normal in the (001) plane and $\sim30^{0}$  relative to the easy axis. (b) The temperature dependence of the average waiting time of 10 switching events for different values of $H$. Inset: the average waiting time $\bar{\tau}$ vs 1/T with a reversing field of 2.8 T. The error bars indicate confidence interval of $90\%$. (c) The temperature dependence of the magnetic field for which the average waiting time $\overline{\tau}$ is closest to 300 s. The error bars indicate the higher and lower field or the higher and lower temperature for which $\overline{\tau}$ was measured.
} \label{trapping}
\end{figure}
The expected thermally activated nature of the magnetization reversal is  most clearly tested by waiting time experiments performed by applying a reversing field which induces reversal within a measurable waiting time $\tau$. Figure 3(a) shows such measurements performed with two different reversing fields.

In the upper two plots we prepare the magnetization in one state and apply a reversing field of ${\rm3.4 \ T}$. We then measure the transverse resistance (${\rm R_{xy}}$) which is sensitive to the perpendicular component of the magnetization as a function of time, and the figure shows the data points until a reversal (manifested in sign change of ${\rm R_{xy}}$) occurs. The experiment is repeated a number of times and for each time the reversal occurs after a different waiting time. The figure shows the distribution of the waiting time until reversal with a reversing field of ${\rm3.4 \ T}$ at ${\rm20.25}$ and ${\rm19 \ K}$ and we see that a change of  ${\rm <10 \%}$ in temperature yields a noticeable change in the distribution of the waiting time.

In the lower two plots of Fig. 3(a) we show the same type of experiments with a reversing field of ${\rm3.725 \ T}$ at ${\rm7}$ and ${\rm3 \ K}$ and we see that despite the much bigger change in temperature, there is no noticeable  change in the distribution of the waiting time for reversal.

Figure 3(b) is a summary of experiments as shown in Fig. 3(a). We show the temperature dependence of the average waiting time ($\bar{\tau}$) for reversal using 4 different fields (which correspond to 4 different energy barriers for reversal). We see that above 10 K the average waiting time increases by an order of magnitude when the temperature is decreased by ${\rm <10 \%}$. It increases monotonically from 78 s at 50 K to 470 s at 48.5 K ($\mu_{0}H$=1.8 T), from 22 s, at 29 K to 220 s, at 27 K ($\mu_{0}H$=2.8 T)(see also the inset of Fig. 3(b)) and from 64 s at 20.25 K to 475 s at 19 K ($\mu_{0}H$=3.4 T). On the other hand, the average waiting time is practically the same below 10 K ($\mu_{0}H$=3.725 T).

Furthermore, above 10 K the temperature dependence of the average waiting time is consistent with the expectation for thermal activation assisted reversal that $\tau= \tau_0 e^{E_b/k_B T}$. This is demonstrated by the fit in the inset of Fig. 3(b) which shows the temperature dependence of the average waiting time with a reversing field of 2.8 T at a temperature interval between 27 and 29 K. The fit yields
$\tau_0 \sim 10^{-13}-5\times10^{-12}$, consistent with the measured ferromagnetic resonance frequency of $\sim200$ GHz \cite{Kerr}. The energy barrier is estimated using $E_{b}\sim \mu_{0}M_sH_c(1-H/H_c)^{\alpha}V$ where we use $M_s=250$ KA/m and $H_{c}$ is the field for which $E_{b}=0$. Based on low-temperature measurements we estimate that $\mu_{0}H_{c}\geq3.75$ T and using $\alpha=1.5$ \cite{alfa} yields a volume $V$ of $<150$ $\rm{nm}^{3}$. This volume is much smaller than the volume of the measured pattern (more than $50 000$ nm$^{3}$), suggesting that the reversal occurs via nucleation followed by propagation.

 Figure 3(c) demonstrates the crossover at 10 K in a different way. It shows the temperature and magnetic field for which the average waiting time $\bar{\tau}$ is closest to 300 s (due to the exponential dependence of $\tau$ on $H$ and $T$ we actually take the field for which log $\bar{\tau}$ is closest to $~\log(300)$). Here too there is a clear crossover to temperature-independent behavior below ${\rm 10 \ K}$.

\begin{figure}[t]
\includegraphics[scale=0.47]{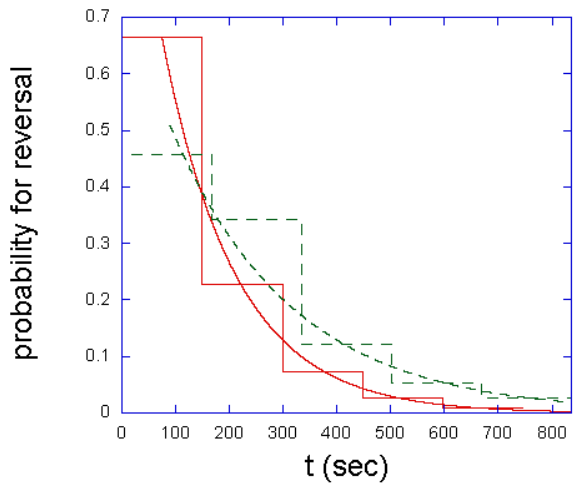}
\caption{The reversal probability for different time intervals at 4 K (red, solid line) and 14 K (green, dashed line) based on hundreds of reversals. The lines are fits to $\int^{t+\Delta/2}_{t-\Delta/2}(1/\bar{\tau})\exp(-t/\bar{\tau})dt$ where $\Delta$ is the chosen time interval for the histograms.} \label{trapping}
\end{figure}

 Although the temperature dependence of $\bar{\tau}$ is strikingly different above and below 10 K, $\bar{\tau}$ exhibits exponential distribution both above and below 10 K and the error bars are calculated accordingly. Figure 4 shows the distribution of $\tau$ at  4 and 14 K
 using hundreds of reversals and compares it with a probability distribution of the form $(1/{\bar{\tau}})\exp(-t/{\bar{\tau}})$.

The striking crossover at such a high temperature calls for special effort to exclude possible artifacts.
 Although the temperature in our commercial measuring system (PPMS 9) is measured by several sensors, we have mounted a temperature sensor
  exactly where we mount our samples and verified the accuracy of our temperature reading. We also excluded possible effects of the probing current by verifying that the observed crossover
 does not depend on the magnitude of the current nor on the rate at which we probe our sample.

 The main remaining question is whether it is plausible to attribute our observation to MQT-dominated magnetization reversal below ${\rm 10 \ K}$. Following Ref.  \cite{MQT_theory_BOOK} a rough estimate for the crossover temperature $T_{c}$, below which the reversal is expected to be dominated by MQT is given by the relation $T_{c}\sim\mu_{B} H_{a}$, which yields $T_{c}\sim5$ K  for $\mu_{0}H_{a}\sim7$ T. To estimate the volume that reverses via MQT we use the expression for the tunneling rate $\Gamma$ given in Ref. \cite{MQT_theory}: $\Gamma=A\exp(-B)$ where $A=([\frac{15}{2\pi}]^{1/2}B^{1/2}\omega_{0})$, $K_{B}T_{c}=\frac{1}{\pi}\hbar\omega_{0}$, where $B=\frac{16\times6^{1/4}}{5}S\epsilon^{5/4}|\cot\theta|^{1/6}$, $\epsilon=(1-\frac{H}{H_{c}})$, $S$ is the total spins, and $\theta$ is the angle between the anisotropy field and the applied magnetic field. For $\epsilon\sim 0.02$, $\theta\sim150^{0}$, $T_{c}\sim10$ K, and $0.1>\Gamma>0.001$ s$^{-1}$ we find that the total $S$ corresponds to a volume smaller than $100$ nm$^{3}$, similar to the upper bound obtained in the thermal activation regime.

In conclusion, we have demonstrated that the magnetization reversal in nanostructures of $\rm{SrRuO_3}$ exhibits a clear crossover from thermally activated reversal to temperature-independent reversal below 10 K. The results raise the intriguing possibility that the low-temperature reversal is dominated by MQT. The correspondence between the high crossover temperature and the large magnitude of the magnetic anisotropy in $\rm{SrRuO_3}$ suggests that systems with higher magnetic anisotropy are good candidates for observing MQT at higher temperatures.

We acknowledge useful discussions with J. S. Dodge, Y. Kats, and E. M. Chudnovsky. L.K. acknowledges support by the Israel Science Foundation founded by the Israel Academy of Sciences and Humanities.  J.W.R. grew
the samples at Stanford University in the laboratory of M. R. Beasley.


\begin{thebibliography}{9}

\bibitem{Neel} L. N\'{e}el, Ann. Geophys. \textbf{5}, 99 (1949).

\bibitem{Brown_pr} W. F. Brown, Phys. Rev. \textbf{130}, 1677 (1963).

\bibitem{Brown_euro} W. Wernsdorfer, E. Bonet Orozco, K. Hasselbach, A. Benoit, B. Barbara, N. Demoncy, A. Loiseau, H. Pascard, and D. Mailly, Phys. Rev. Lett. \textbf{78}, 1791 (1997).

\bibitem{Cullity} B. D. Cullity and C. D. Graham, $Introduction to Magnetic Materials$, 2nd ed. (Wiley, New York, 2009), p.384.

\bibitem{MQT_theory_BOOK} E. M. Chudnovsky and J. Tejada, Macroscopic Quantum Tunneling of the Magnetic Moment (Cambridge University Press, Cambridge, 1998), p.122.

\bibitem{MQT_theory} M.-Carmen Miguel and E. M. Chudnovsky, Phys. Rev. B \textbf{54}, 388 (1996).

\bibitem{Mn} J. R. Friedman, M. P. Sarachik, J. Tejada, and R. Ziolo, Phys. Rev. Lett. \textbf{76}, 3830 (1996).

\bibitem{Fe} C. Sangregorio, T. Ohm, C. Paulsen, R. Sessoli, and D. Gatteschi, Phys. Rev. Lett. \textbf{78}, 4645 (1997).

\bibitem{Ni} K. Hong and N. Giordano, J. Magn. Magn. Mater. \textbf{151}, 396 (1995).

\bibitem{ErAs} F. Coppinger, J. Genoe, D. K. Maude, X. Kleber, L. B. Rigal, U. Gennser, J. C. Portal, K. E. Singer, P. Rutter, T. Taskin, A. R. Peaker, and A. C. Wright, Phys. Rev. B \textbf{57}, 7182 (1998).

\bibitem{BaFe} W. Wernsdorfer, E. Bonet Orozco, K. Hasselbach, A. Benoit, D. Mailly, O. Kubo, H. Nakano and B. Barbara, Phys. Rev. Lett. \textbf{79}, 4014 (1997).

\bibitem{srruo} G. Koster, L. Klein, W. Siemons, G. Rijnders, J. S. Dodge, C.-B Eom, D. H. A. Blank, and M. R. Beasley, Rev. Mod. Phys. \textbf{84}, 253 (2012).

\bibitem{Ms} L. Klein, J. S. Dodge, C. H. Ahn, J. W. Reiner, L. Mieville, T. H. Geballe, M. R. Beasley and A. Kapitulnik, J. Phys.: Condens. Matter \textbf{8}, 10111 (1996).

\bibitem{Kerr} M. C. Langner, C. L. S. Kantner, Y. H. Chu, L. M. Martin, P. Yu, J. Seidel, R. Ramesh, and J. Orenstein, Phys. Rev. Lett. \textbf{102}, 177601 (2009).

\bibitem{EA_Kats} Y. Kats, I. Genish, L. Klein, J. W. Reiner and M. R. Beasley, Phys. Rev. B \textbf{71}, 100403 (2005).

\bibitem{Domain strips} A. F. Marshall, L. Klein, J. S. Dodge, C. H. Ahn, J. W. Reiner, L. Mieville, L. Antagonazza, A. Kapitulnik, T. H. Geballe, and M. R. Beasley, J. Appl. Phys. \textbf{85}, 4131 (1999).

\bibitem{wall_width} I. Asulin, O. Yuli, G. Koren, and O. Millo, Phys. Rev. B \textbf{74}, 092501 (2006).

\bibitem{reciprocity_theorem} M. B$\rm \ddot{u}$ttiker, IBM J. Res. Dev. \textbf{32}, 317 (1988).

\bibitem{alfa} R. H. Victora, Phys. Rev. Lett. \textbf{63}, 457 (1989).

\end{thebibliography}
\end{document}